\documentclass[prl,aps,showpacs,twocolumn,amsmath,superscriptaddress]{revtex4}
\usepackage[dvips]{color}
\usepackage{graphicx} 
\usepackage{bm}

\newcommand{\be}{\begin{equation}}
\newcommand{\ee}{\end{equation}}
\newcommand{\bea}{\begin{eqnarray}}
\newcommand{\eea}{\end{eqnarray}}
\newcommand{\bse}{\begin{subequations}}
\newcommand{\ese}{\end{subequations}}
\renewcommand{\d}{\mathrm{d}}
\renewcommand{\a}{a_{\mathrm{1D}}}
\newcommand{\chio}{\chi^{(0)}}
\newcommand{\pf}{k_{\mathrm F}}
\newcommand{\kf}{k_{\mathrm F}}
\newcommand{\ef}{\varepsilon_{\mathrm F}}
\newcommand{\im}{\mathrm{Im}}
\newcommand{\gb}{g_\mathrm{B}}
\newcommand{\gf}{g_\mathrm{F}}
\newcommand{\q}{q}

\newcommand{\eprint}[1]{#1}

\begin{document}
\title{The Dynamic Structure Factor of the
1D {B}ose Gas near the {T}onks-{G}irardeau Limit}

\author{Joachim Brand}
\affiliation{Max Planck Institute for the Physics of Complex
Systems, N\"othnitzer Stra{\ss}e 38, 01187 Dresden, Germany}

\author{Alexander Yu.~Cherny}
\affiliation{Max Planck Institute for the Physics of Complex
Systems, N\"othnitzer Stra{\ss}e 38, 01187 Dresden, Germany}
\affiliation{Bogoliubov Laboratory of Theoretical Physics, Joint
Institute for Nuclear Research, 141980, Dubna, Moscow region, Russia}
\date{\today}

\begin{abstract}

While the 1D Bose gas appears to exhibit superfluid response under
certain conditions, it fails the Landau criterion according to the
elementary excitation spectrum calculated by Lieb.  The apparent
riddle is solved by calculating the dynamic structure factor of the
Lieb-Liniger 1D Bose gas.  A pseudopotential Hamiltonian in the
fermionic representation is used to derive a Hartree-Fock operator,
which turns out to be well-behaved and local. The Random-Phase
approximation for the dynamic structure factor based on this
derivation is calculated analytically and is expected to be valid at
least up to first order in $1/\gamma$, where $\gamma$ is the
dimensionless interaction strength of the model. The dynamic structure
factor in this approximation clearly indicates a crossover behavior
from the non-superfluid Tonks to the superfluid weakly-interacting
regime, which should be observable by Bragg scattering in current
experiments.
\end{abstract}

\maketitle

\begin{figure}
\includegraphics[width=\columnwidth]{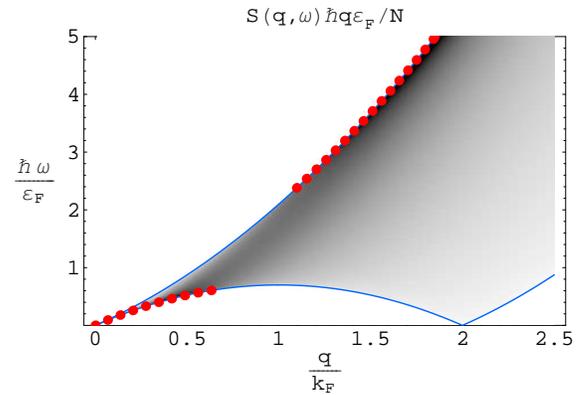}
\caption{\label{fig:DSFgamma20} Excitation spectrum at $\gamma = 13$.
The upper and lower thin (blue) lines show the dispersions
$\omega_+(q)$ and $\omega_-(q)$ of Eq.~(\ref{eqn:dispersions}),
respectively, limiting the elementary excitations of the LL model. The
dynamic structure factor $S(q,\omega)$ in the approximation (\ref{eqn:strdyn})
is shown in shades of grey, between zero (white) and 1.0 $
N/(\hbar q \ef)$ (black). Here $\kf=\pi n$ and $\ef=\hbar^2 \kf^2 /(2
m)$. The the presence of a
$\delta$-function contribution is indicated by red dots.}
\end{figure}

The emergence of superfluidity at low temperatures is one of the most
dramatic manifestations of quantum many-body physics in
nature. Although the gas of interacting Bosons in 1D is described by
the exactly solvable Lieb-Liniger (LL) model \cite{lieb63:1,lieb63:2},
the question of superfluidity is not readily answered from the exact
solutions. The dimensionless interaction parameter 
$\gamma = \gb m /(\hbar^2 n)$ \cite{lieb63:1} governs the crossover
from the weakly-interacting quasicondensate for $\gamma \ll 1$ to the
strongly-interacting Tonks-Girardeau gas \cite{girardeau60} at
$\gamma=\infty$, where $\gb$ is the coupling constant of the 1D Bose
gas 
\cite{note:coupling},
$n$ is the line density, and $m$ is the particle mass.  The Landau
criterion of superfluidity \cite{pitaevskii03:book} predicts
a critical velocity $v_{\text{c}}$ for the breakdown of
dissipationless flow if no elementary excitations of momentum $p$ are
accessible with energy below $p v_{\text{c}}$ to dissipate its
energy. The excitation spectrum of the LL model \cite{lieb63:2}
contains umklapp excitations at finite momentum $2\pi n$ and energies which
tend to zero in a large system predicting a critical velocity of zero
by the Landau criterion {\it for any value of $\gamma$} [see the
lower thin (blue) line in Fig.~\ref{fig:DSFgamma20}]. On the other hand,
Luttinger-liquid theory and instanton techniques~\cite{kagan00,buchler:100403}
predict superfluidity in the LL model for sufficiently small $\gamma$.
A resolution of this apparent
paradox lies in the probability of excitation by an infinitesimal
external perturber which is given by the dynamic structure factor (DSF)
$S(q,\omega)$, the Fourier transform of the time-dependent
density-density correlation function. 
A suppression of transitions in the vicinity of the um\-klapp excitation
was found in Refs.~\cite{sonin71,castro_neto94,pitaevskii04}.
In this Letter, we
report a calculation of the DSF,
valid for large $\gamma$ as
shown in Fig.~\ref{fig:DSFgamma20}.
We find that even at fairly large but finite $\gamma$ the low energy
umklapp excitations are indeed suppressed while
higher-energy excitations are enhanced. This is to be
contrasted with the Tonks-Girardeau gas at $\gamma=\infty$ where all
excitations are equally well accessible and prevent superfluidity.

Experiments have recently made significant progress in confining Bose
gases in 1D~\cite{bongs01}
probing the strongly-correlated Tonks-Girardeau regime by increasing
interactions up to values of $\gamma\approx 5.5$ \cite{Kinoshita2004}
and to effective values of $\gamma_{\text{eff}}\approx 200$ in an
optical lattice \cite{paredes04}.
In a recent
experiment~\cite{stoferle:130403}, the zero momentum
excitations of a 1D Bose gas in an optical lattice have been measured
by Bragg scattering, a technique that could also be used to measure the
DSF.

Although the LL model is exactly solvable, it is notoriously difficult to
calculate the various correlation functions. Many results in limiting
cases are summarized in the book~\cite{korepin93}, but the full
problem is not yet 
solved. Recently, progress has been made on various time-independent
correlation functions \cite{Gangardt2003a,olshanii02ep}.
While a wealth of information is available for small $\gamma$ where
Bogoliubov's perturbation theory can be applied, the
strongly-interacting regime was hardly accessible as a systematic
expansion in $\gamma^{-1}$ was lacking.

In this Letter we study the repulsively interacting 1D Bose gas in a
fermionic representation, introduced as follows.
The Bose-Fermi mapping of Girardeau \cite{girardeau60} can be generalized~\cite{cheon99} to
map the 1D Bose gas with pair interactions $V(x)=\gb
\delta(x)$ to an equivalent {\it interacting} Fermi system with the same
excitation energies and the same absolute values of the associated wave functions.
Hence,
any observable has the same value in both representations if it
derives from a local operator in coordinate space, {\it i.e.}~is a
function of the density operator $\hat{n}(x)=\sum_{i} \delta(x-x_{i})$.
In particular,
this is the case for the DSF $S(q,\omega)$, which is the
Fourier transform of the density-density 
correlation function~\cite{pines89,pitaevskii03:book}, and expresses the
probability to excite a particular excited state through a density
perturbation 
\begin{equation}
S(q,\omega)=
\sum_{n}|\langle0|\hat{\rho}_{q}|n\rangle|^{2}\delta(\hbar
\omega-E_{n}+E_{0}) , 
\label{sqomega}
\end{equation}
where $\hat{\rho}_{q}=\sum_i \exp(-i q x_i)$ is the Fourier component
of the density operator. 
In this mapping, large interactions between bosons correspond to
small interactions between fermions, which allows
for simple variational or perturbative
treatment. However, the short-range interaction of Cheon and Shigehara~\cite{cheon99}
is highly singular for this purpose.
It has been pointed out by Sen~\cite{sen99}
that the ground state energy functional can be derived variationally with 
the pair-interaction pseudopotential
\be \label{eqn:SenPP}
  V_{\rm Sen}(x_1,x_2) =
  -\gf \delta''(x_1-x_2),
\ee
where $\gf = {2 \hbar^4}/({m^2 \gb})$ is the coupling constant in
the fermionic representation. This pseudopotential,
however, is applicable only  if the 
variational functions are continuous and vanish whenever two particle
coordinates coincide. This is the case for Slater determinants which
we will use later to derive the Hartree-Fock (HF) and Random-Phase
approximations (RPA) but not for
the exact fermionic wavefunctions \cite{cheon99}.
$V_{\rm Sen}$ thus generates the first order
correction to the  
ground-state energy easily
whereas Sen found a renormalizable divergence in second order
perturbation theory.

Recently, a more general pseudopotential was proposed by
Girardeau and Olshanii  \cite{girardeau03ep},
which is
also defined for 
discontinuous functions and generates the correct energy
functional for arbitrary values of the coupling constant. We give the
following useful representation of this 
pseudopotential in terms of an integral kernel:
\begin{align}
\nonumber 
&V(x_1,x_2;x'_1,x'_2)= \\& -{2}\gf \delta\left(
\frac{x_1+x_2-x'_1-x'_2}{2} \right) 
\delta'(x_1-x_2)\delta'(x'_1-x'_2). 
\label{eqn:our}
\end{align}

Based on a pseudopotential Hamiltonian (\ref{eqn:SenPP}) or (\ref{eqn:our}), we
can derive the HF approximation and the generalized
RPA giving access to the density-density
correlation function and the static and dynamic structure factors. The
results should be valid at least to first order in $\gamma^{-1}$ and thus be
useful near the Tonks-Girardeau limit of $\gamma =\infty$, where the
interacting boson problem maps exactly to a free Fermi gas
\cite{girardeau60}.
The purpose of this work is thus twofold: Besides predicting the behavior of
correlation functions of the 1D Bose gas, we also test the validity
and usefulness of the fermionic pseudopotential approach within the
RPA.


The HF approximation for the fermionic system is derived 
variationally
in the standard way. We find that both pseudopotentials
(\ref{eqn:SenPP}) and (\ref{eqn:our}) lead to exactly the same
result: The HF equations for the single-particle orbitals $\varphi_j(x)$
take the usual form
\begin{equation}\label{eqn:HF}
\hat{F}\varphi_j(x)=\varepsilon_j\varphi_j(x).
\end{equation}
Unlike the case of Coulomb interactions where the Fock operator
$\hat{F}$ is nonlocal with a local Hartree and a nonlocal exchange
term we find a purely local Fock operator
\begin{align} \label{eqn:HFoperator}
\hat{F}=&-\frac{\hbar^2}{2m}\frac{\partial^2}{\partial x^2}+V_{\rm
ext}(x)\nonumber\\
&+
\gf \left[n(x)\frac{\partial^2}{\partial x^2}+
{2}{\cal P}(x)i\frac{\partial}{\partial x}
-{\cal T}(x)\right].
\end{align}
The first two terms on the right hand side are just the
single-particle part of the Hamiltonian. The mean-field parts in the
square bracket involve the single-particle density $n(x) = \sum_j n_j
{{\varphi}^{*}}_{j}(x)\varphi_{j}(x)$ and the derivative densities ${\cal
  P}(x)=-i\sum_j n_j {{\varphi}^{*}}'_{j}(x)\varphi_{j}(x)$
and ${\cal T}(x)=\sum_j n_j
[{\varphi}^{*}_{j}(x)\varphi''_{j}(x) +
  2{{\varphi}^{*}}'_{j}(x)\varphi'_{j}(x)]$, reminiscent of momentum
and energy densities, respectively.
Here, $\varphi' = \d \varphi/\d x$ and $n_j$ is the fermionic
occupation number
$n_j={1}/\{\exp[(\varepsilon_j-\mu)/(k_{\mathrm B} T)]+1\}$.
At temperature $T=0$, $n_j$ defines the Fermi sea.

For the homogeneous gas ($V_{\rm ext}\equiv 0$), the quantum number
$j$ is associated with the particle momentum, and the 
single-particle orbital solutions of Eq.~(\ref{eqn:HF}) are plane waves with
energy and effective mass
\begin{align} \label{eqn:epsHF}
\varepsilon_q=&{\hbar^2 q^2}/{(2m^*)}-\gf {\pi^2} n^3/{3},\\
m^* =& m /\left(1-{2 \gf m n}/{\hbar^2}\right) = {m}/({1-4
   \gamma^{-1}}), 
\end{align}
respectively. The HF approximation for the ground-state energy 
coincides with the first two terms of the large-$\gamma$
expansion of the exact ground-state energy in the LL model
\cite{lieb63:1}. 

The HF approximation permits us to calculate the linear response
function of time-dependent HF, also known as RPA with exchange or
generalized RPA \cite{pines89}. 
We calculate the DSF at zero temperature by the relation \cite{pines89}
$S(q,\omega)= \im \{\chi(q,-\omega-i\varepsilon) / \pi\}$ for $\omega>0$ and
$S(q,\omega)=0$ for $\omega<0$ from the dynamic polarizability
$\chi(q,\omega+i\varepsilon)$. The function $\chi(q,\omega+i\varepsilon)$ determines the linear
density response to an external field \cite{pines89,pitaevskii03:book}
and can be obtained 
from the linearized equation of motion of the density operator in the
time-dependent HF approximation. In the Fourier representation of
momentum and frequency,
algebraic equations result which can be solved for
$\chi(q,\omega)$.
In the thermodynamic limit,
the sums over momenta
are replaced by integrals.
We find $\chi(q, z) = {\chio(q, z)} /
\{(1-4\gamma^{-1}) [B-D\chio(q,z)]\}$, with $z=\omega+i\varepsilon$ and 
$\varepsilon {\to} +0$. 
For the DSF this yields for $\omega>0$
\begin{align}
  S(q,\omega)=&
    \frac{- \chi^{(0)}_{2}(q,\omega) B} {\pi (1-4\,\gamma^{-1})
    \left[\Big(B-D\chi_{1}^{(0)}\Big)^2 + 
    \Big(D\chi_{2}^{(0)}\Big)^2\right]}\nonumber\\
  & + \delta[\omega - \omega_0(q)] A(q) ,
\label{eqn:strdyn}
\end{align}
where
\begin{align}
B=&1-{4\,(3\,\gamma -16)}/{(\gamma-4)^3} ,\\
  D=&\frac{4 \ef}{N}\frac{\gamma}{(\gamma -4)^2}\Bigg\{\frac{q^2}{\pf^2}
  \frac{2\gamma - 9} {2 \gamma}
  - \frac{2}{\gamma}\\& -
 \left[\frac{\hbar (\omega+i\varepsilon)\, \pf}{
  \ef q} \right]^2 \frac{3\gamma -16}{2( \gamma  -4)^2}\Bigg\} ,
\end{align}
with the Tonks gas' Fermi wavenumber $\kf=\pi n$ and 
energy $\ef =  \hbar^2 \kf^2/(2 m)$. 
The polarizability $\chio = \chio_1 +i \chio_2$ of the ideal 1D Fermi
gas with renormalized mass.
is given by the real and imaginary parts
\begin{align}
  \chi^{(0)}_{1}(q,\omega)=&\frac{N m^{*}}{2\hbar^2 q \kf}\ln\left|
  \frac{\omega^{2} - \omega^{2}_{-}(q)}{\omega^{2}_{+}(q)-\omega^{2}}
  \right|, \label{chi1}\\
  \chi^{(0)}_{2}(q,\omega)=&- \frac{N \pi m^{*}}{2\hbar^2 q \kf}\left\{
  \begin{array}{ll} 
  \pm 1,& \omega_{-}\le \pm \omega\le \omega_{+},\\[2mm]
  0,& \text{else,}
\end{array}\right.
\label{chi2}
\end{align}
where the dispersion relations
\be \label{eqn:dispersions}
  \omega_\pm(q) = {\hbar |2 \pf q \pm q^2|}/{(2 m^*)}
\ee
border the 
continuum part of the accessible excitation spectrum made up from HF
quasiparticle-quasihole excitations~(\ref{eqn:epsHF}), see Fig.~\ref{fig:DSFgamma20}.
The $\delta$-function part of the DSF (\ref{eqn:strdyn}) relating to
discrete excitations of collective character lies outside of the
continuum part and comes from possible zeros in the denominator of
$\chi(q,z)$. It is determined by the solution $\omega_0(q)$ of
the transcendental equation $B-D\chi_{1}^{(0)}=0$. We have solved this
equation in various limits and found that at most one solution for
$\omega_0(q)$ may exist. The strength $A(q)$ is given by the residue
of $\chi$ at the pole $z=\omega_0(q)-i\varepsilon$. For small
$\gamma^{-1}$, the strength $A(q) \simeq 2 N \gamma \exp(-\gamma\q/\kf)$
is exponentially suppressed and possible solutions are close to the
dispersions $|\omega_0-\omega_\pm| \propto
\exp(-\gamma\q/\kf)$. Due to this proximity of the discrete and
continuous parts and expected smearing of discrete contributions
beyond RPA, we may conjecture that the
$\delta$-function should be seen as part of the continuum, enhancing
contributions near the border. In fact, we know from the exact
solutions that energy 
spectrum is continuous \cite{lieb63:2}.

At finite $\gamma$ we find a $\delta$-function
contribution below and close to $\omega_-$ for small $q$ whereas for
large $q$ there is a discrete contribution at energies larger than
$\omega_+$, see Fig.~\ref{fig:DSFgamma20}. In the limit $\q \to \infty$ at finite $\gamma$, the
$\delta$ part completely determines the DSF as the continuum part
vanishes, and asymptotically  $A\simeq N$, and  $\omega_0 \simeq \hbar
q^2/(2 m)$ becomes 
the free particle dispersion reminiscent of the DSF for the weakly
interacting Bose gas in Bogoliubov theory \cite{pitaevskii03:book}. A
more detailed analysis of the $\delta$-function part will be given
elsewhere \cite{cherny04tbp}.

The branches $\omega_\pm$ as shown in Fig.~\ref{fig:DSFgamma20} are
large $\gamma$ approximations for the type I and II elementary
excitation branches, respectively, introduced by Lieb
\cite{lieb63:2}. In accordance with the exact results, both branches
share the same slope at the origin and give rise to a single speed of
sound given by 
$ v_s \equiv {\d \omega_\pm}/{\d k} = {\hbar \kf}/{m^*}$, which is the
correct first order expansion of $v_s$ for large $\gamma$
\cite{lieb63:2}. Note that the usual Bogoliubov perturbation theory
for weakly interacting Bosons gives a similar expansion of $v_s$ for
small $\gamma$ and the type I excitation branch. Type II excitations
are not described in the Bogoliubov theory.

\begin{figure}
\includegraphics[width=\columnwidth]{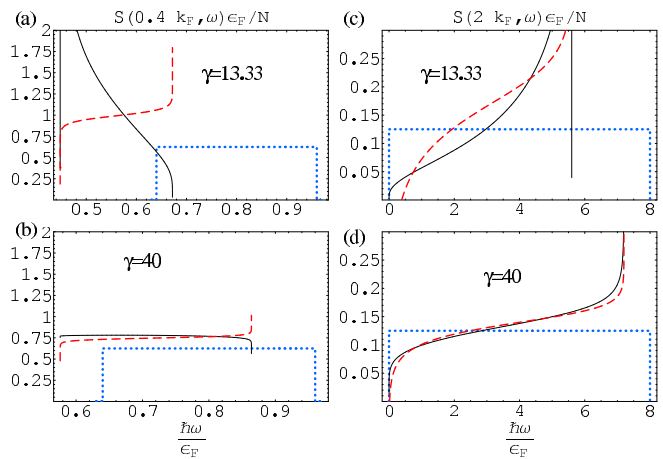}
\caption{\label{fig:DSFk2} DSF $S(q,\omega)$ as a function of
$\omega$ at $q=0.4 \kf$ [(a) and (b)] and $q=2 \kf$ [(c) and (d)],
and at interaction strength $\gamma=13.33$ [(a) and (c)] and
$\gamma=40$ [(b) and (d)]. The solid (black) line shows the RPA
result~(\ref{eqn:strdyn}), the dashed (red) line shows the first order
expansion in $\gamma^{-1}$~(\ref{eqn:first-oder}), and the dotted
(blue) line shows the DSF of the Tonks limit ($\gamma=\infty$) for
comparison. While the 
first-order expansion always has a divergence to $+\infty$ near
$\omega_+$, the RPA result shows an enhancement of low-energy
excitation near $\omega_-$ at small $\q$. Note the unphysical
negative values of the DSF in the first order expansion near
$\omega_-$, in particular for the umklapp excitations in panel (c).}
\end{figure}
 
Due to a logarithmic
singularity in $\chi_{1}^{(0)}$, the DSF vanishes on the dispersion
curves. For the Tonks gas at $\gamma\to\infty$, the value of the
DSF within these limits is independent of $\omega$ and takes the value
of $N m/(2 \pi \hbar^2 q n)$.
The energy-dependence in the
RPA for finite $\gamma^{-1}$ is shown in
Figs.~\ref{fig:DSFgamma20} and \ref{fig:DSFk2}. In particular, we see
that the umklapp excitations at $q=2 \kf$ at
small $\omega$, prohibiting superfluidity of the Tonks gas, are being
suppressed. We find that $S(2\kf,\omega)$ in the RPA approaches zero as
$1/\ln^2(\hbar\omega/\ef)$, in contrast to the results~\cite{castro_neto94,pitaevskii04},
predicting a power-law dependence on
$\omega$ for finite $\gamma$ based on a pseudoparticle-operator approach.

The RPA result may be expanded in $1/\gamma$,
which yields
\be\label{eqn:first-oder}
  S(q,\omega)\frac{\ef}{N} = \kf \frac{1+ 8\gamma^{-1}}{4 q} +
   \frac{\ln 
  f(q,\omega)}{2\gamma} + {\cal O}(\gamma^{-2})
\ee
with $f(q,\omega) = |(\omega^2 - \omega_-^2)/(\omega_+^2 - \omega^2)|$.
This expansion is supposed to be consistent with first order
perturbation theory. However, it can assume negative values as seen in
Fig.~\ref{fig:DSFk2} although $S(q,\omega)$ is known to be strictly
non-negative, a property that is fulfilled by our RPA result
(\ref{eqn:strdyn}). Close to $\omega_+$, the first order expansion has
has a logarithmic singularity to $+\infty$ which may be a precursor of
the dominance of Bogoliubov-like excitations in the DSF at small
$\gamma$. In the RPA this effect is even more pronounced due to a
strong and narrow peak of the DSF in the RPA near $\omega_+$ at large
momenta as seen in Fig.~\ref{fig:DSFk2}. At finite gamma and for small
momenta, however, the RPA predicts a peak near $\omega_-$, contrary
to the first-order result. Whether this effect is real or an artefact
of the RPA is not obvious and may be decided by more accurate
calculations or experiments. Spurious higher order terms in the RPA
and an improved approximation scheme have been discussed in
Ref.~\cite{brand98}. On the other hand, Roth and Burnett have recently
observed a qualitatively similar effect in numerical calculations of
the DSF of the Bose-Hubbard model \cite{roth04}.

We have verified numerically that the $f$-sum rule $m_1\equiv \hbar^2
\int \omega 
S(q,\omega) \d \omega = N \hbar^2 q^2/(2 m)$ is fulfilled, which
should be an exact statement for the RPA from a general theorem
\cite{thouless61}. The same follows also from the the large $\omega$
asymptotics of $\chi(q,\omega)\simeq 2m_{1}/(\hbar \omega)^2$ when $\chi$ is
analytic as a function of $\omega$ in the upper half complex plane
\cite{pitaevskii03:book}. Our approximation breaks down when $\gamma\lesssim 8$
for small values of momenta $q\lesssim \pf$,
 where imaginary poles of $\chi$ appear.

Our result (\ref{eqn:strdyn}) defines approximations for the static
structure factor $S(q) = ({\hbar}/{N}) \int S(q,\omega) \d \omega$ and
the pair correlation function $g(x) = 1+\int [S(q) - 1]/(2 \pi n)
\exp(iqx) \d x$ that will be discussed in detail elsewhere
\cite{cherny04tbp}. Here we only note that the approximation for
$S(q)$ is continuous and has the low-momentum expansion $S(q) = q [1+4 \gamma^{-1}-
q^2/(\kf^2 \gamma)]/(2 \kf) +{\cal O}(q^5)/\gamma +{\cal
O}(\gamma^{-2})$. We also find that $g(x=0)$ vanishes in first order
of $\gamma^{-1}$ as consistent with Ref.~\cite{Gangardt2003a},
indicating once more the validity of our results.

Summarizing, we have derived variational approximations that should
yield valid expansions of various correlation functions of the 1D Bose
gas in $1/\gamma$, making predictions in a so-far unexplored
regime. These approximations prove consistent with known limits and
sum rules and therefore establish the usefulness of the fermionic
pseudopotentials (\ref{eqn:SenPP}) and (\ref{eqn:our}). We have also
shown a rare example where a fully analytical calculation of the
generalized RPA can be carried out for a non-trivial many-body system.
Our results have immediate relevance to current experimental
endeavors to explore fermionization in the strongly-interacting 1D
Bose gas as they indicate qualitative deviation in the DSF from the
Tonks gas limit already for very small $\gamma^{-1}$. The more
accessible region of intermediate values of $\gamma$, which would
yield more insight into the superfluid properties, is beyond the scope
of this study. Answers may come from higher order or
density-functional-theory based approximations extending the work of
Ref.~\cite{brand04a}, from numerical calculations, or from experimental
measurements. Obvious extensions of the present study to finite
temperature, inhomogeneous systems, and periodic lattice potentials are under way.

The authors acknowledge discussions with Sandro Stringari who inspired
their interest in this problem.


\end{document}